\documentclass[aps,prl,floatfix,epsfig,twocolumn,showpacs,preprintnumbers]{revtex4}
\usepackage{amsmath,amssymb,graphicx,bm,epsfig}

\newcommand{\eps}{\epsilon}

\newcommand{\vk}{{\mathbf{k}}}
\newcommand{\vK}{{\mathbf{K}}}
\newcommand{\vq}{{\mathbf{q}}}
\newcommand{\vQ}{{\mathbf{Q}}}
\newcommand{\vPhi}{{\mathbf{\Phi}}}
\newcommand{\vS}{{\mathbf{S}}}
\newcommand{\cG}{{\cal G}}
\newcommand{\cF}{{\cal F}}
\newcommand{\cD}{{\cal D}}
\newcommand{\Tr}{\mathrm{Tr}}

\begin{document}

\title{Optical Conductivity of the t-J model within Cluster Dynamical Mean Field Theory}
\author{Kristjan Haule and Gabriel Kotliar}
\affiliation{Department of Physics, Rutgers University, Piscataway, NJ 08854, USA}
\date{\today}

\begin{abstract}
We study the evolution of the optical conductivity in the t-J model
with temperature and doping using the Extended
Dynamical Cluster Approximation. The cluster approach results in an
optical mass which is doping independent near half filling.  The
transition to the superconducting state in the overdoped regime is
characterized by a decrease in the hole kinetic energy, in contrast to
the underdoped side where kinetic energy of holes increases upon
superfluid condensation. In both regimes, the optical conductivity
displays anomalous transfers of spectral weight over a broad frequency
region.
\end{abstract}
\date{\today}
\maketitle

How superconductivity emerges from a strongly correlated normal state
is one of the most important problems in the field of strongly
correlated electron systems. This problem has received intensive
attention of the past two decades spurred by the discovery of high
temperature superconductivity. A large number of experimental probes
have been applied to this problem, but there is, at this point, no
theoretical concensus on the basic physics or the minimal model
required to describe this rich phenomena.

In recent years, significant progress has been achieved through the
development of the Dynamical Mean Field Theory (DMFT)
\cite{old_review} and its generalizations
\cite{DCA,CDMFT,Lichtenstein_cluster}. This approach, allows us to
isolate which aspects of the physics of a given material can be
understood within a local approach, and what is the minimal cluster
size required to describe certain phenomena.  For example it has been
shown that a single site method describes well the physics of the
finite temperature Mott transition in materials such as $\mathrm{V_2
O_3}$ in a broad region of temperature and pressures around the Mott
endpoint, while the interplay of Peierls and Mott instabilities in
materials such as $\mathrm{V O_2}$ and $\mathrm{Ti_2 O_3}$ requires a
two site link as a minimal reference frame \cite{Biermann}.  There is
an evidence that the Hubbard or the t-J model treated by a four site
cluster DMFT approach captures many essential aspects of the physics
of the cuprates such as its phase diagram
\cite{Maier_review,DCA,Maier_EDCA}, the variation of the spectral
function and the electron lifetime along the Fermi surface
\cite{Civelli,Parcollet,Civelli_SC}. This approach systematizes and
extends the early slave boson treatments of the cuprates which now
appear as restricted low energy parameterizations in a more general
approach.

One of the most powerful bulk probes of carrier dynamics is
optical conductivity. Previous studies of optics in t-J and Hubbard model
\cite{Jarrelloptics,Prelovsek} have
allowed an accurate description of many experimentally observed
anomalies found in the cuprate superconductors at high temperatures,
around or above the normal temperature. In this letter we use the cluster
extension of DMFT, which allows us to study the evolution of the
optical conductivity in the most interesting temperature regime around
the transition from anomalous metal to d-wave superconductor.
As a result, we gain several new insights and we show theoretically
that: a) while optical spectral weight is reduced in the underdoped
side, the optical mass of the carriers remains constant. b) The
changes in hole kinetic energy when entering the superconducting state
are much smaller than in the Hubbard model, and are positive in the
overdoped side and negative in the underdoped regime. c) The broad
redistribution of spectral weight which is observed in the
superconducting state is largely due to the anomalous Greens function,
and in this respect optical measurements combined with theory allows
us to extract additional information not available from photoemission
spectroscopy. d) At optimal doping, the low frequency part of optical
conductivity is Drude like while the intermediate range shows
approximate powerlaw $\omega^{-2/3}$.  All these findings are
consistent with experiments and can be understood in terms of a
plaquette embedded in a correlated medium but not in terms of a single
site theory.

\textit{Formalism}
Our starting point is the t-J model which was proposed by P.W.
Anderson as a minimal model to describe the cuprate
superconductors. We use an exact reformulation of this model in
terms of a spin fermion model using the  Hubbard-Stratonovich
transformation to decouple the spin interaction. The effective action
describing the interaction of fermions with spin
fluctuations takes the following form
\begin{widetext}
\begin{eqnarray}
  S=\int_0^\beta d\tau\left\{ \sum_{\vk\sigma}c_{\vk\sigma}^\dagger(\tau)
  (\frac{\partial}{\partial\tau}-\mu+\eps_\vk)c_{\vk\sigma}(\tau)
  +\sum_i U n_{i\uparrow}(\tau)n_{i\downarrow}(\tau)
  +\sum_\vq\left[{\vPhi^\dagger}_\vq (\tau)\frac{2g^2}{J_\vq}\vPhi_\vq(\tau)
  +ig\; \vS_\vq\left(\vPhi_\vq^\dagger(\tau)+\vPhi_{-\vq}(\tau)\right)\right]
  \right\}
\end{eqnarray}
\end{widetext}
The Hubbard U term in the action will be taken to infinity to enforce
the constraint.

The exact Baym-Kadanoff-like functional for this problem is
\begin{eqnarray}
  \Gamma[\cG,\cD]&=&-\Tr\log(G_{0}^{-1}-\Sigma)-\Tr[\cG\Sigma]\nonumber\\
  &+&\frac{1}{2}\Tr\log(\cD_{0}^{-1}-\Pi)
  +\frac{1}{2}\Tr[\cD\Pi]+\Phi[\cG,\cD].
  \label{functional}
\end{eqnarray}

We chose Extended Dynamical Cluster Approximation method (EDCA)
\cite{DCA,Maier_review,Maier_EDCA,mojPRL} on a plaquette because it is
the simplest mean field theory that justifies ignoring the vertex
corrections of optical conductivity.  The only approximation of the
EDCA is to replace the Green's functions in the interacting part of
the functional $\Phi[\cG,\cD]$ with the corresponding course grained cluster Green's
functions $\cG_\vk\rightarrow\cG_\vK=\sum_{\vk\in\vK}G_\vk$ and
$\cD_\vq\rightarrow\cD_\vQ=\sum_{\vq\in\vQ}\cD_\vQ$
where the sum $\sum_{\vk\in\vK}$ is over those $\vk$
momenta in Brillouin zone which correspond to certain cluster momenta
$\vK$ (see \cite{Maier_review}).
The saddle point equations of the functional
Eq.~\ref{functional} are $\Sigma_\vK = {\delta\Phi}/{\delta\cG_\vK}$
and $\Pi_\vQ = 2{\delta\Phi}/{\delta\cD_\vQ}$ which automatically give
pice-vise constant self-energies. Together with the Dyson equations
$\cG_\vK=\sum_{\vk\in\vK}(G_0^{-1}-\Sigma_\vK)^{-1}$ and
$\cD_\vQ=\sum_{\vq\in\vQ}(D_0^{-1}-\Pi_\vQ)^{-1}$ form a closed set of
equations. Few comments are in order: i) The bosonic self energy is
simply related to the spin susceptibility \cite{my-phd} $\chi_\vq=
(g^2\Pi_\vQ^{-1}+J_\vq)^{-1}$ and this expression can be used to
calculate bosonic self-energy knowing local spin-susceptibility.  ii)
The Baym-Kadanoff functional $\Phi[\cG,\cD]$ which depends only on the
cluster Green's functions can be obtained by solving the cluster
problem coupled to the fermionic and bosonic bath
\cite{mojPRL,Maier_EDCA}.
As an impurity solver we use exact diagonalization for the cluster and
an NCA approach to treat the hybridization of the cluster with the
bath.  To study superconductivity we allow the off-diagonal long range
order by employing the Nambu method of anomalous Green's functions.

The optical conductivity of the plaquette within EDCA is
particularly simple because the vertex corrections vanish just
like in single site DMFT. The reason is that all four cluster
$\vK$-points of the plaquette are course-grained over the part of
the Brillouin zone which is particularly symmetric, namely $\vk$
as well as $-\vk$ point for each $\vk$ point are in the same
cluster region. Since the current vertex depends only on the
cluster $\vK$ point, the vertex legs can be closed and
coarse-grained $\sum_{\vk\in\vK} v_\vk \cG_\vk* \cG_\vk$. Since
velocity is even function of $\vk$ and $\cG_\vk$ at fixed cluster
$\vK$ depends of $\vk$ only through $\eps_\vk$, this quantity
vanishes.

The optical conductivity can therefore be simply expressed by
\begin{eqnarray}
  \sigma'(\omega) &=& \sum_{\vk \sigma}
  e^2 v_\vk^2\; \int\frac{dx}{\pi}
  \frac{f(x-\omega)-f(x)}{\omega}\times \nonumber\\
  && \left[\cG_{\vk}^{''}(x-\omega)\cG_{\vk}^{''}(x)
  +\cF_{\vk}^{''}(x-\omega)\cF_{\vk}^{''}(x)
  \right]
\label{opt_conductivity}
\end{eqnarray}

\textit{Results} - 
\begin{figure}
\includegraphics[width=0.85\linewidth]{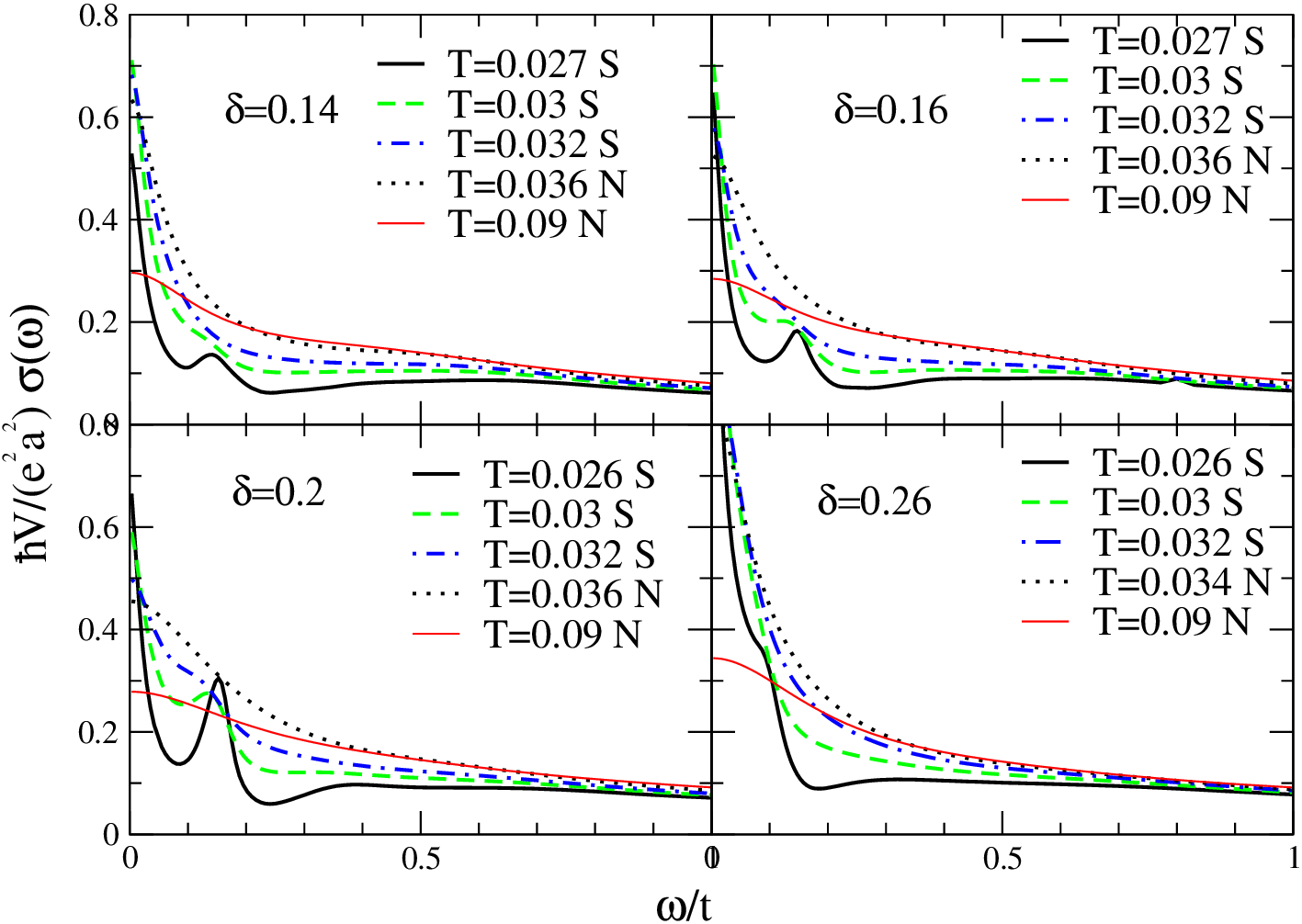}
\caption{
  Optical conductivity across the superconducting transition for
  various doping levels from underdoped to overdoped regime. Letter S
  in legend stand for superconducting state and N for normal state of
  the t-J model.
}
\label{optics_linear}
\end{figure}
In Fig.~\ref{optics_linear} we show the evolution of  the optical
conductivity  for various temperatures and dopings from room
temperature $T\approx 0.09t$ down to transition temperature
$T_c\approx 0.034 t$ and slightly below the transition $T \approx
0.76 T_c$. We have representatives of the  three regimes of the
cuprate superconductors, slightly underdoped $\delta=0.14$,
optimally doped $\delta=0.16-0.2$ and the overdoped case
$\delta=0.26$.  In agreement with experiments on high Tc's, and
with earlier theoretical single site DMFT studies, the optical
conductivity extends over a broad frequency range, and consist of
a broad  Drude component at low frequencies that sharpens with
decreasing temperature  and a higher frequency ("mid infrared")
component with  substantial intensity at high frequencies.

Upon entering the superconducting state no real optical gap opens
below the transition, unlike the standard case of the s-wave BCS
superconductors. However, a substantial reduction of the optical
conductivity due to superconductivity is observed up to very high
frequency even beyond $\omega > t \approx 2500 \mathrm{cm}^{-1}$ which
is ten times bigger that the superconducting gap at optimal doping
$\Delta\approx 0.1t$ itself.

The Drude-like low frequency conductivity narrows as the temperature
decreases from room temperature to just above $T_c$ and continues to
grow even below the transition temperature.  This enhancement of the
low frequency conductivity as a result of the onset of coherence is
seen in many experimental studies \cite{opticsw} and should be
contrasted with the reduction of the conductivity at higher
frequencies due to the opening of a gap.

\begin{figure}[ht]
\includegraphics[width=0.9\linewidth]{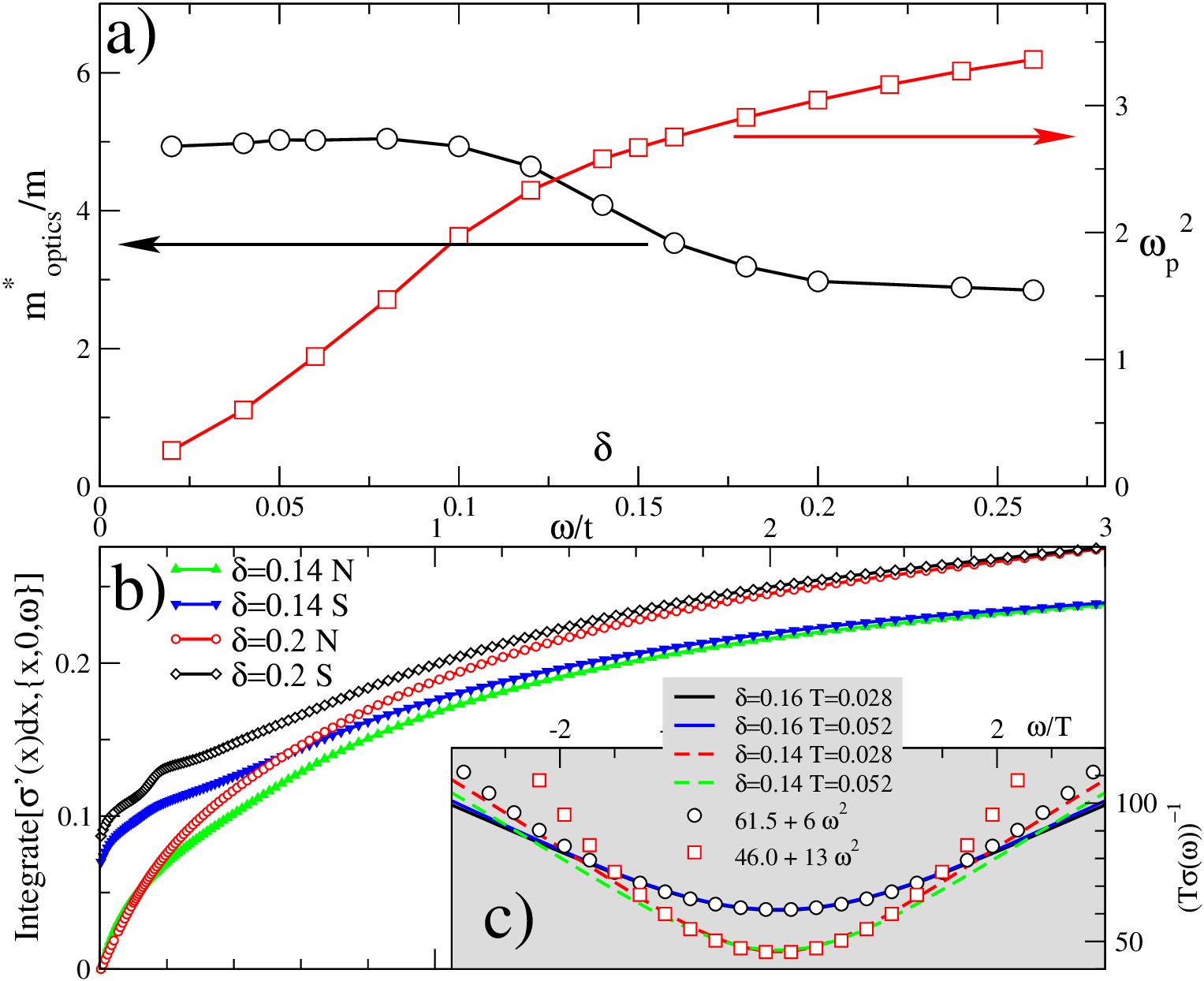}
\caption{
  a) Optical effective mass as determined from Eq.~\ref{effe_mass} and
  plasma frequency versus doping. The plasma frequency decreases
  towards the Mott point because of decrease of the active part of the
  Fermi surface while effective mass remains almost constant.
  b) The integral of the optical conductivity $\int_0^\omega\sigma'(x)dx$
  for normal and superconducting state. The optical weight in SC state
  missing in the finite frequency region collapses into $\delta$
  function. The spectral weight is restored when integral is taken up
  to $\sim 4t\sim 1eV\approx 8000cm^{-1}$.
  c) The inverse of low frequency optical conductivity
  $1/(T\sigma(\omega))$ versus $\omega/T$ is quadratic parabola
  independent of temperature in the optimal doping regime. Optical
  conductivity is thus Drude-like at low frequency up to $\omega\sim
  T$.
} \label{eff_mass}
\end{figure}

The optical conductivity has a very long tail of incoherent spectral
weight that dominates the optics. Furthermore, the tail seems to have
an approximate power law $\sigma\propto (-i\omega)^{\alpha}$ with the
exponent close to $\alpha=2/3$ \cite{preparation} in agreement with
experiments \cite{Bontemps_powerlaw,VanDerMarelScience}. At low
frequency, optical conductivity is Lorentzian-like at optimal doping
with $\omega/T$ scaling. In Fig.~\ref{eff_mass}c we plot
$(T\sigma(\omega))^{-1}$ as a function of $\omega/T$ which is temperature
independent quadratic parabola in the region $-T\lesssim\omega\lesssim
T$ in agreement with recent experiments \cite{VanDerMarelScience}.

The peak around $0.15t$ is the standard BCS coherence peak arising
from the excitations across the superconducting gap in one particle
density of state and is not
visible in experiment which is closer to the clean limit than our
theoretical calculation which overstimates the scattering rate.

Optical conductivity  measurements can be used to define an optical
effective mass  defined by
\begin{equation}
  \frac{m^*}{m}=\lim_{\omega\rightarrow 0}\frac{\omega_p^2}{4\pi\omega}\frac{\sigma''(\omega)}{\sigma''(\omega)^2+\sigma'(\omega)^2}
  \label{effe_mass}
\end{equation}
The plasma frequency in this equation is proportional to the kinetic
energy of the model and at the same time, to the integral of the
optical conductivity.  This definition can be motivated by the
extended Drude model of reference \onlinecite{ExDrMo}. Here we see
that the cluster DMFT method introduces new physics not captured by
the single site approach.

In single site DMFT the optical mass diverges as the Mott transition
is approached. Taking into account the low frequency and low
temperature expansion of the self-energy in the local Fermi-liquid
state $\Sigma(\omega)=\mu-\mu_0+\omega(1-\frac{1}{Z})-i\Gamma$ one
obtains for the conductivity
$\sigma(\omega)=e^2\Phi_{xx}(\mu_0)(1+i\omega/(2 Z\Gamma))/(2\Gamma)$
and therefore the optical mass in single site DMFT is $m^*/m=1/Z$ and
diverges as Mott transition is approached since $Z$ vanishes.

In the cluster case, the mass no longer diverges. In
Fig.~\ref{eff_mass}a, we show the optics mass from Eq.~\ref{effe_mass}
where $\omega_p$ was computed by integrating the optical conductivity
up to $4t\approx 1eV$. When approaching Mott transition, the effective
mass becomes 5-times bigger than the electron mass while it drops down
to 3-times the electron mass in the overdoped regime. This behavior is
very similar to the recent experiments on cuprates reported in
\onlinecite{eff_massc}. The plasma frequency in Fig.~\ref{effe_mass},
which is proportional to the integral of optical conductivity, grows
linearly at small doping and the slope gets less steep around 10\%
doping in accordance with recent experiments on cuprates
\cite{plasma_frequency}.

The EDCA equations in the $2\times 2$ plaquette provide us with a
simple description of how the the Mott insulating state is approached,
since one has only four different patches in momentum space, which
however, change very differently with doping.  The self energy that is
relevant to the optical conductivity corresponds to the patch around
$(0,\pi)$ and $(\pi,0)$ and has a quasiparticle residue which is
approximately doping independent and finite ($Z\approx 0.1$) as the
Mott insulator is approached. However, the optical conductivity
vanishes with vanishing doping just like the plasma frequency in
Fig.~\ref{eff_mass}. This effect is driven by the vanishing of the
number of carriers measured by the length of the active segment of the
Fermi surface, which goes to zero as the Mott point is approached,
rather than with vanishing of quasiparticle renormalization
amplitude. This interpretation of the EDCA results is consistent with
the results obtained by the cumulant periodization of CDMFT
\cite{Tudor}.

{\it Transfer of spectral weight} - A surprising aspect of the physics
of strongly correlated materials, is that low energy phenomena affect
the spectra of the material over a very large energy scale.  This
general phenomena is illustrated in Fig.~\ref{eff_mass}b, which shows
the integral of optical spectral weight in the normal and the
superconducting state.  Low energy phenomena like the onset of
superconductivity which involves a scale of a fraction of J, involves
redistribution of optical weight of the order of $4t \approx 1eV$
which is 40 times more than the gap value. This large range of
redistribution of spectral weight was also measured on cuprates and
pointed out in \cite{Bontemps_redistribution,Molegraf}.  A new
theoretical insight is that the high frequency redistribution of
weight comes from the anomalous Greens function $\cF*\cF$ in
Eq.~(\ref{opt_conductivity}) and hence can not be observed in the
density of states or ARPES measurements.

{\it Sum Rule} - To understand the mechanism of high temperature
superconductivity, much effort was recently devoted to measure the
change of kinetic energy upon superfluid condensation
\cite{Bontemps_kinetic,Molegraf}. Experimentally this is measured by
integrating conductivity up to large enough cutoff of the order of
$1eV$ in both normal and superconducting state, and interpreting the
results in terms of the optical sum rule which states
\begin{equation}
 \int_0^\Lambda \sigma'(x)dx = -\frac{\pi e^2}{4}\langle T\rangle
\label{sumri}
\end{equation}
This sum rule has also been extensively discussed in the context of
the Hubbard model, in which case, assuming nearest neighbor hopping
matrix elements only, the operator on the right hand side of
Eq.~\ref{sumri} is the kinetic energy of the electrons in the Hubbard
model. It has been shown by Jarrell and collaborators
\cite{Jarrell_kinete} that entering the superconducting state results
in a reduction of kinetic energy, in stark contrast with the BCS model
where the kinetic energy increases upon entering the superconducting
state.

The kinetic energy of the Hubbard model is composed of two different
contributions. The superexchange energy of the spins, and the kinetic
energy of the holes.  In the Hubbard model the optical transitions at
energies of order U, give rise to the superexchange energy of the
spins, while the low energy transitions correspond to the kinetic
energy of the holes, but it is not possible to separate those two
contributions in the Hubbard model.  Here we sharpen the analysis of
Ref.~\onlinecite{Jarrell_kinete}, by evaluating separately these two
physically different contributions, which allows us to make direct
contact with optical experiments. Namely, the experiments measure the
kinetic energy of the holes since the cuttoffs which are used are such
that the transitions into the upper Hubbard band are not included.

\begin{figure}[ht]
\includegraphics[width=0.8\linewidth]{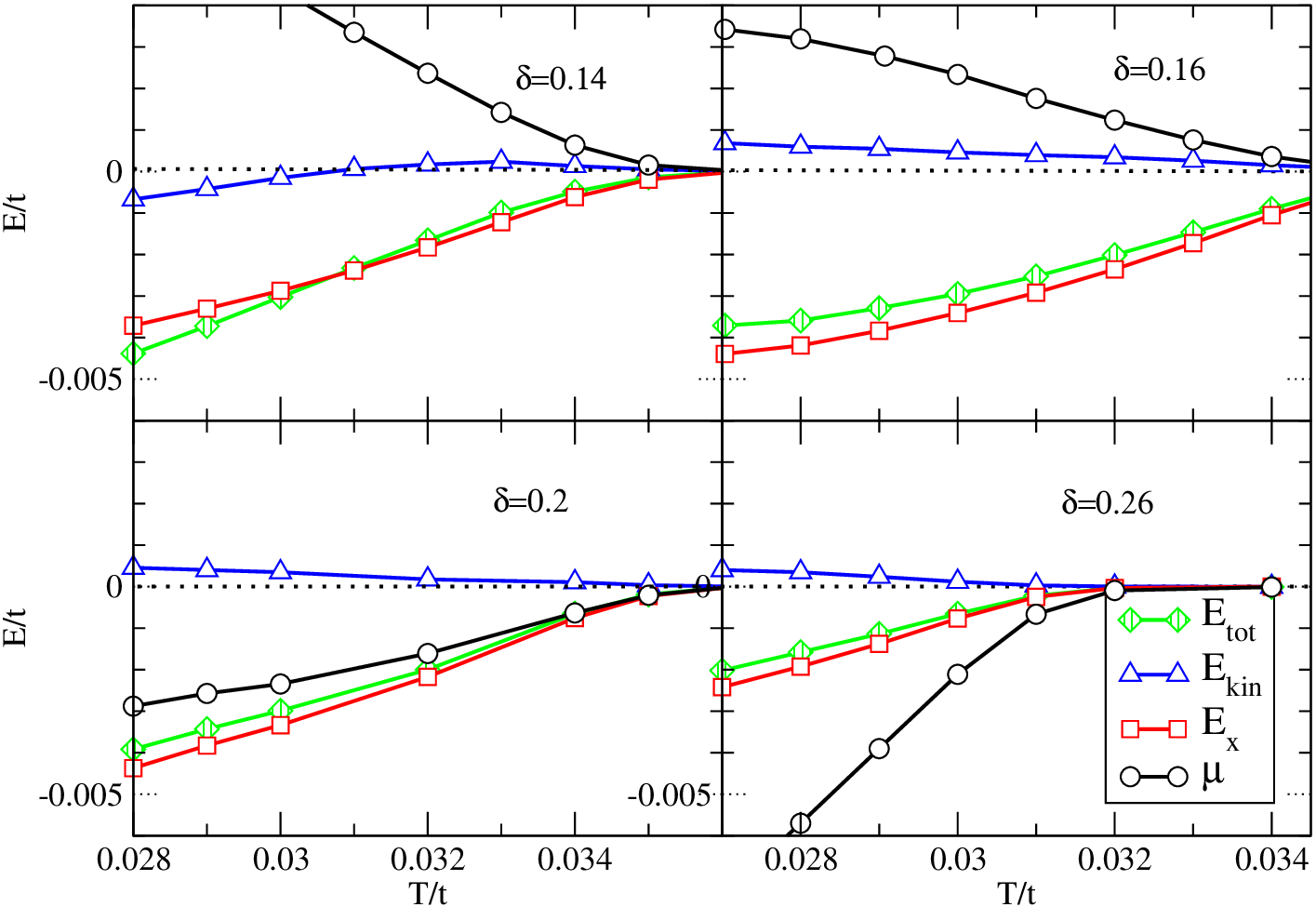}
\caption{
  The difference between the superconducting and normal state energies
  as a function of temperature.
  The following curves are shown:
  blue with triangles up - $E_{kin-S}-E_{kin-N}$;
  red squares - $E_{exch-S}-E_{exch-N}$;
  green diamonds - $E_{tot-S}-E_{tot-N}$;
  black with circles - $\mu_S-\mu_N$.
} \label{Energies}
\end{figure}
In Fig.~\ref{Energies}, we show the change of kinetic energy of
holes, the superexchange energy ($\sum_{ij}J_{ij} \langle\vS_i
\vS_j\rangle/2$) and total energy upon condensation as obtained
from our EDCA cluster calculation. The kinetic energy can be
directly calculated from the spectral function and the
superexchange energy from the spin susceptibility.
In agreement with very recent experiments on cuprates
\cite{Bontemps_kinetic} and in agreement with RVB and slave-boson
prediction, we see that the change in kinetic energy upon superfluid
condensation changes sign between overdoped to underdoped regime. In
overdoped regime, the conventional BCS picture is applicable where
kinetic energy of holes increases while the superexchange energy
decreases just like the interaction energy in conventional phonon
mediated superconductors. The later change is much larger such that
the total condensation energy is positive. In underdoped regime, the
superexchange energy still decreases upon condensation and gives the
largest contribution to the condensation energy
\cite{Maier_EDCA}. However, the kinetic energy of holes is gained in
uderdoped case so that the kinetic energy gives positive contributes
to condensation in agreement with recent experiment
\cite{Bontemps_kinetic}.

In Fig~(\ref{Energies}), we also show the change of chemical potential
upon condensation. Similarly to kinetic energy, it also changes sign
with doping and it increases in underdoped regime while it decreases
in overdoped regime. We believe this is a consequence of the asymmetry
of the normal as well as the superconducting one particle density of
states.

\textit{Conclusion} - In conclusion, we have studied the t-J model
using the EDCA in a plaquette near the optimally doped regime. The
theoretical results capture the main features of the optical
conductivity near the transition from a strongly correlated metal to a
superconductor. In spite of the shortcomings of the approximation used
(the EDCA $2\times 2$ cluster overestimates the superconducting
critical temperature and the quasiparticle scattering rates) we
believe that the lessons from this mean field theory, as to the
physical origin of the scaling of the optical mass with doping, the
anomalous redistribution of spectral weight and the changes of kinetic
and exchange energy upon entering the superconducting state are
general, and can be derived with considerable more effort in other
cluster schemes, or using larger clusters. They stress the importance
of the differentiation of states in momentum space in strongly
correlated materials near the Mott transition.

\section{Acknowledgments}
We are grateful to Nicole Bontemps, Antoine Georges and Peter W\"{o}lfle for useful
discussions. GK was supported by NSF DMR Grants No. 0528969.

\end{document}